# Monotonic and cyclic mechanical reliability of metallization lines on polymer substrates

Oleksandr Glushko[a], Andreas Klug[b,1], Emil J.W. List-Kratochvil[b,c], Megan J. Cordill[a*]

[a] Erich Schmid Institute of Materials Science, Austrian Academy of Sciences and Dept. Material Physics, Montanuniversität Leoben, Jahnstrasse 12, A-8700 Leoben, Austria, oleksandr.glushko@oeaw.ac.at, megan.cordill@oeaw.ac.at

[b] NanoTecCenter Weiz Forschungsgesellschaft mbH, Franz-Pichler-Straße 32, A-8160 Weiz, Austria

[c] Institut für Physik, Institut für Chemie & IRIS Adlershof, Humboldt-Universität zu Berlin, Brook-Taylor-Straße 6, 12489 Berlin, Germany, emil.list-kratochvil@hu-berlin.de

*Corresponding Author: Megan J. Cordill, megan.cordill@oeaw.ac.at

**Abstract**

Mechanical stability of Ag and Cu printed and evaporated metallization lines on polymer substrates is investigated by means of monotonic tensile and cyclic bending tests. It is shown that lines which demonstrate good performance during monotonic tests fail at lower strains during a cyclic bending tests. Evaporated lines with the grain size of several hundreds of nanometers have good ductility and consequently good stability during monotonic loading but at the same time they fail at low strains during cyclic bending. Printed lines with nanocrystalline microstructure, in contrast, demonstrate more intensive cracking during monotonic loading but higher failure strains during cyclic bending. Apart from the grain size effect, the effect of film thickness on the saturation crack density after cyclic bending is also demonstrated. Thinner films have higher crack density in accordance with the shear lag model.

---

[1] Present Address: AVL List GmbH, Hans-List-Platz 1, A-8020 Graz, Austria, andreas.klug@avl.com

# 1. Introduction

Electronics developed for flexible applications, such as rollable displays, wearable sensors, circuitry or flexible batteries [1–5], need to have the ability to stretch and bend without failure. This attribute is in contrast to rigid electronic counterparts which mostly fail through electrical degradation or thermal-mechanical instabilities, such as electromigration, crack formation, or interface delamination. Much research has gone into determining failure mechanisms and fatigue lifetimes of materials used in flexible electronics. Methods include in-situ tensile straining to study crack and deformation evolution [6–8], stress-strain development with X-ray diffraction (XRD) [8–11], and interface adhesion measurements [12–14]. The electrical behavior can also be examined with tensile straining with two- or four-point-probe resistance geometry incorporated into the grips [15–17] or with wires attached to the sample [18,19] for tensile straining. The advantage of in-situ resistance measurements is the precise measurement of the crack onset strain (COS), or failure strain, of brittle films and lines, such as transparent conductive oxides or brittle metals. The COS can be considered the threshold for stretching of flexible technologies. Not only the COS of brittle materials can be determined with in-situ resistance measurements during straining, but also the electrical degradation with continued strain and any resistance recovery during the unloading of the strain [16,20]. Both the COS and electrical degradation can be used to evaluate the influence of thickness, microstructure, and deposition method.

Bending techniques have also been developed to address the rollable and bending applications and are necessary to prove flexibility [21–26]. The use of bending tests also better simulates real-life loading conditions. The bending tests provide information about the strain below which no damage or fracture occurs and when performed cyclically, a lifetime curve can be determined by the number of cycles the system can withstand at a given strain before damage is observed. Bending strain at which failure occurs is important for flex-to-connect applications where the device is only bent once, while cyclic bending is necessary for rollable or foldable applications when many bending cycles are expected. The development of reliability curves from bending tests provide important information about the lifetime of metal and ceramic films on flexible substrates [22]. However, different research groups have different definitions of the number of cycles to failure for films on substrates depending on the detection method used. When resistance is the detection parameter, then the number of cycles to failure is determined when the relative resistance increases 2% [26] or even 20-25% [25,27]. Other definitions of the number of cycles to failure include when the extrusion density reaches saturation [28–30] or simply using the first cycle where damage is observed [31].

Many studies have addressed the role of film thickness [25,32,33], the annealing treatments of printed Ag films [26,34–36], and the fabrication method (printing versus evaporation) [26,34], however, the combined role of film thickness and film microstructure has not been thoroughly addressed for monotonic and cyclic bending tests. In this paper, Ag and Cu metallization lines with different thicknesses and microstructures are characterized by means of monotonic tensile straining and cyclic bending tests in combination with in-situ resistance measurements and post-mortem electron microscopy. It is shown that high ductility crucial for good reliability during a monotonic tensile test may act as a drawback limiting the mechanical stability under cyclic loading. Further examination of the mechanical behavior demonstrates that the grain size and film thickness are the main length scale parameters which strongly influence the mechanical performance of conductive lines.

## 2. Materials and Methods

An overview of six different sample types, namely evaporated and printed silver as well as evaporated copper lines, considered in this paper are listed in Table I. All lines have the width of 0.5 mm and the length of 40 mm. The evaporated silver lines were produced in a vacuum coating unit using a base pressure less than 1x10$^{-6}$ mbar and a deposition rate of 10 Å/s. Three different nominal thicknesses were deposited, 100, 400, and 800 nm on 130 µm thick PEN (Polyethylene Naphthalene, Teonex® Q65HA) substrates using a silver source (purity 99.95%) without additional substrate heating. For comparison, Cu lines with the thickness of 800 nm were also evaporated onto 180 µm thick PET (Polyethylene Terephthalate, Melinex® ST506) substrates using the same vacuum coating unit and shadow mask. For the inkjet printed lines, a PiXDRO LP50 printer equipped with FujiFilm Dimatix Spectra S-Class SE-128 piezoelectric print head was utilized with silver ink CCI-300 from Cabot Corporation. During printing the substrate temperature was kept constant at 60°C to obtain a print resolution of 800 dpi. Samples were sintered at 150°C for 2 hrs in a drying oven after printing. The printed lines were fabricated on the same PEN and PET substrates as the evaporated lines and have a nominal thickness of 800 nm. Initial resistances of all lines measured using the distance between the contacts of 39 mm are given in the Table I. Since all lines have the same width one can conclude that the resistivity of the printed lines is about an order of magnitude higher than the resistivity of evaporated lines which is common for printed Ag [26,34]. The grain sizes of evaporated lines were determined using electron-backscatter diffraction mapping (EBSD). The particle size of the printed lines were taken from the ink specifications and confirmed by direct scanning electron microscopy (SEM) imaging to be between 30-80 nm.

In-situ tensile straining with four-point-probe resistance measurements were performed on an MTS Tyron 250® Universal testing machine using a constant displacement rate of 5 µm/s for loading to 20% strain and unloading to zero force. For each material system, at least four tests were performed with the probing contacts incorporated into the grips and measured with a Keithley 2000 multimeter as described in more detail in [15,16]. After straining, samples were examined with SEM and the average saturation crack densities after 20% strain were determined with a linear line intercept method.

The bending tests were performed using a newly developed apparatus called FLEX-E-TEST [22]. The samples are clamped at one end in grips which are fixed to a motor-driven wheel. Each grip has a curvature radius corresponding to the required bending radius to be applied during the test. During rotation the samples sequentially come into contact with the bending surface and bend to the corresponding grip radius (Figure 1). Three different bending radii of 5 mm, 10 mm, and 20 mm were used in bending experiments. The bending strain, $\varepsilon$, was estimated using the general formula $\varepsilon=d_s/2r$, where $d_s$ is the substrate thickness and $r$ is the bending radius of the grip. Note that the PET substrates are slightly thicker than the PEN substrates (Table I) which results in higher strains under the same bending radius. In order to analyze the evolution of the damage in the lines a series of tests were performed with different numbers of bending cycles between 100 and $5\times10^4$. For each maximum cycle number three different samples were tested in order to exclude the effects of possible fabrication defects. The amount of mechanical damage induced by cyclic bending was characterized in terms of the linear crack density. The crack density was calculated by taking a series of SEM micrographs after a given number of bending cycles. For each micrograph the number of cracks per unit length was determined and then averaged.

**3. Results**

3.1. Monotonic tensile straining

An overview of the behavior of the electrical resistance during monotonic tensile straining to a maximum strain of 20% is shown in Figure 2a for the evaporated lines and in Figure 2b for printed lines. Each curve has a loading portion which starts at zero strain, increasing to the maximum strain of 20% and an unloading portion, as shown by the arrows in Figure 2. Following the common and well-justified practice [15,37], the constant volume approximation ($R/R_o = (L/L_o)^2$, where $R$ is the measured resistance, $R_o$ is the initial resistance, $L$ is the measured gauge length and $L_o$ is the initial gauge length), which shows the resistance behavior in the case of perfect plastic deformation without cracking (dashed "theory" curve), was used as a reference curve. The point where the experimentally measured relative resistance ($R/R_o$) starts to deviates from the constant volume approximation curve defines the value of the COS. The 400 nm and 800 nm thick evaporated silver lines as well as the 800 nm thick evaporated copper lines demonstrate very good mechanical stability (Figure 2a) because the resistance follows the constant value approximation curve up to the strain of approximately 12%. Furthermore, at the maximum strain of 20% the deviation from the theoretical curve is moderate compared to the 100 nm thick evaporated silver lines. Such resistance behavior suggests that the applied strain induced only short, isolated cracks or localized deformation. The resistance of the 100 nm thick evaporated silver line starts to deviate from the theoretical curve after only 5% strain and reaches a higher value at the maximum strain (120% resistance increase). Faster growth of resistance with decreasing film thickness was previously observed [25,37] and is explained by the lower ductility of the thinner films, most likely due to the smaller grain size.

The resistance of the printed silver lines grows by several hundred percent as the strain is increased to 20% (Figure 2b). Such a strong deviation from the theoretical curve indicates that extensive cracking occurred during straining. Indeed, since the size of silver particles used for printing lie well below 100 nm, plastic deformation through dislocation slip [38] is prohibited, resulting in rather brittle behavior. The values of COS for the printed lines are in the range of 5%. It is necessary to note, that for many flexible electronics applications which do not involve elastomeric substrates, there is no need to prove high stability at such high strains as 20%. Much more important is good reliability at low strains below 3%. This consideration is based on the fact that polymer substrates such as PEN, PET, and polyimide start to deform plastically for strains above 3% and should not be used in device applications requiring higher strains. Thus, for this lower strain regime printed Ag lines on PEN and PET show very good stability comparable to that of evaporated Ag lines.

Another important aspect of the in-situ resistance experiments is that all of the relative resistance curves shown in Figure 2 demonstrate resistance recovery during unloading. The solid circles depict the resistance of fully unloaded samples when the applied load has reached zero. As it was shown in [16] this resistance recovery is caused by both the mechanical relaxation of the substrate and crack closure. The thicker Ag and Cu evaporated lines almost completely recover indicating extensive crack re-bridging. This behavior is often overlooked but is highly important to accurately predict the real electrical behavior.

The differences in the in-situ resistance behavior discussed above are also manifested in the morphology of the mechanical damage induced in the lines during tensile straining. Figure 3 presents a comparison of the damage of three evaporated silver lines of different thickness, the printed Ag lines on PEN and PET, as well as the evaporated Cu line is presented. Note that all micrographs are at the same magnification. The 100 nm thick evaporated Ag lines (Figure 3a) exhibit clear channel cracks with lengths up to several micrometers and an average crack density of $0.33 \pm 0.03$ $\mu m^{-1}$ after 20% strain (Figure 3a). This observation explains the faster growth of the electrical resistance of 100 nm thick evaporated lines (Figure 2a). The evaporated 400 and 800 nm thick Ag lines exhibit very short cracks and locally thinned areas indicating that plastic deformation occurred during straining (Figure 3b-c). Since concurrent plastic deformation and cracking occurred, an accurate crack density is difficult to determine with SEM micrographs. However, if only channel cracks are considered for the 400 nm evaporated film, then the average crack density was determined to be $0.13 \pm 0.02$ $\mu m^{-1}$ after 20% strain. The decrease of the saturation crack density with increasing films thickness is expected when comparing the same material and deposition method. The printed Ag lines on PEN and PET (Figure 3d-e) have clear channel cracks, some of which have re-connected due to the removal of the applied load. After 20% strain printed lines revealed an average crack density of $0.22 \pm 0.04$ $\mu m^{-1}$ on PEN and $0.12 \pm 0.03$ $\mu m^{-1}$ on PET. The difference is most likely due to the nominal thickness variation of the substrate or is an indication of the interface shear strength. The 800 nm evaporated Cu lines reveal no cracking but areas of localized deformation (Figure 3f).

3.2. Cyclic bending reliability

For applications of repeated bending the recently developed FLEX-E-TEST technique [22] was used. The damage induced in the conductive lines was characterized by the linear crack density. For simplicity and clarity only the effect of tensile bending cycles will be considered below. The comparison of tensile, compressive and mixed bending is given in detail elsewhere [22].

The dependence of the crack density on the cycle number for the evaporated silver lines with three different thicknesses is shown in Figures 4a and 4b for 5 mm and 10 mm bending radii, respectively. In the case of the 5 mm bending radius (1.3% strain), the crack density saturates after a few hundred cycles (approx. 500) for all three thicknesses. The crack density, however, depends strongly on the film thickness, with the largest values found for 100 nm thick lines and the lowest for 800 nm thick lines. In the case of the 10 mm bending radius (0.65% strain), the crack density of the 100 nm thick lines does not start to increase until after $1 \times 10^4$ cycles and has the lowest value compared to the 400 nm and 800 nm thick lines after $5 \times 10^4$ cycles. Due to the small grain size (100-250 nm) the dislocation-based slip mechanisms are restricted in the 100 nm thick lines and the applied strain of 0.65% is too low to induce intensive homogeneous cracking. A similar behavior was observed in 200 nm evaporated Cu films subjected to 1.1% bending strain [25]. In contrast, dislocations in the 400 and 800 nm thick lines can move at very small strains and accumulate damage quickly with the cycle number because of the larger grain size and increased thickness.

The SEM micrographs of evaporated and printed lines after $5\times10^4$ cycles with 1.3% bending strain are shown in Figure 5. All micrographs are taken at the same magnification. The effect of film thickness is clearly manifested in Figures 5a-c where the 100 nm thick lines demonstrate long channel cracks running parallel to each other. Such fracture morphology is typical for rather brittle films although localized plastic deformation in the form of surface extrusions can be seen along the cracks. The 800 nm thick evaporated Ag lines (Figure 5c) exhibit significant plasticity, which is manifested by numerous slip bands and large extrusions along the cracks. Evidence of plastic deformation was not observed in the printed lines (Figure 5d-e) where the film surface between the cracks is not deformed by cyclic loading as observed in the evaporated lines. The 800 nm evaporated Cu lines (Figure 5f) behave in a similar manner as the 800 nm evaporated Ag lines illustrating extrusions and cracks. The saturation crack densities of the 100 nm and 400 nm evaporated Ag films after $5\times10^4$ cycles of 1.3% bending strain relate well to the monotonic crack densities after 20%. At lower bending strains, however, the values strongly deviate, with the corresponding reasons still under investigation.

The concept of fatigue life, which is defined as the number of cycles to failure, is difficult to apply when characterizing the fatigue failure of polymer-supported films. The main reason is that the films do not completely fracture due to the polymer substrate. This makes a clear and universal definition of failure impossible, but, as mentioned already, several groups have attempted to define failure criteria [26,27,31,39]. Since the cyclic bending method used here does not measure resistance as a function of bending, the following failure criteria is proposed. From the point of view of flexible electronic applications one can define the minimum number of bending cycles which a device must sustain during the entire expected lifetime. Assuming that a device should be able to be bent ten times per day for the lifetime of 10 years, a critical number of 36,500 cycles is obtained. Taking into account possible defects and irregularities during the fabrication process, a more conservative value of $5\times10^4$ cycles seems to be adequate. The cyclic reliability of a flexible component can then be visualized by plotting the dependence of a reliability parameter, such as the electrical resistance or crack density, as a function of the applied strain or bending radius after $5\times10^4$ cycles. Such dependencies of the crack density after $5\times10^4$ cycles on the applied bending strain are shown in Figure 6, where the different printed and evaporated lines with the same thickness of 800 nm on PET and PEN are compared. The most distinct feature of the reliability curves is the existence of a threshold strain for the printed lines below which the crack density is zero. This threshold strain is estimated to be at least 0.4% for printed lines, which is much lower than the COS value (5%) obtained from monotonic tensile straining (Figure 2b). In the case of the 800 nm thick evaporated lines the crack density is non-zero even at the lowest applied strains of 0.33% for both copper and silver making the difference to the COS (12%) even more drastic. What the reliability curves also illustrate is, that despite the monotonic behavior, printed lines are better than evaporated lines at bending strains less than 0.6%. One should also notice that the evaporated Ag lines and the printed Ag on PEN have very similar crack densities at strains between 0.6% and 1.3% despite significant differences in their microstructures. This observation suggests that the saturation crack density is predominantly defined by the film thickness as is well known for monotonic testing using the shear lag theory [40–42].

## 4. Discussion

The presented results illustrate that there is a significant gap between the mechanical reliability assessment obtained from monotonic tensile tests and cyclic bending tests. In the case of monotonic tensile tests the ductility of a thin film is the crucial property which determines the electro-mechanical performance. The more ductile the film, which is controlled by the resulting microstructure, the closer the electrical response is to the constant volume approximation and fewer cracks are induced during straining as demonstrated in Figure 3. High ductility, however, acts as a drawback under cyclic loading conditions. It is known that in ductile metals dislocations can move at very low strains, typically below 0.1%. During a single bending cycle dislocations can be activated and accumulated with increasing cycles, leading to the formation of extrusions and cracks even at low cyclic strains (Figure 5). The printed lines and, to some extent, the 100 nm thick evaporated lines have restricted ductility due to the small grain or particle size. This is clearly demonstrated by the lower COS values deduced from a monotonic tensile test (Figure 2). At the same time, if the applied mechanical strains remain below the threshold value, then no damage is induced even after $5 \times 10^4$ bending cycles. This threshold strain corresponds to the true elastic limit stress below which the deformation is purely elastic. The exact value of the elastic strain limit is difficult to determine since it depends on the fabrication technique, thickness, and resulting microstructure, but also on the density of defects which can act as crack-initiation sites.

Apart from the grain size effect on the reliability there is a strong effect of film thickness on the saturation crack density (Figure 4a). The dependence of the saturation crack density on film thickness can be understood with the help of a shear lag model [41–43]. Formation of a crack in a thin film on a substrate results in stress relaxation in the immediate vicinity of the crack. Due to the stress relaxation the next crack can appear only at some distance away from the existing crack. This distance, which corresponds to the minimum crack spacing, was shown to depend directly on the film thickness being smaller for thinner films and larger for thicker films [40-43].

Although the shear lag model, in general, considers intrinsically brittle films under increasing monotonic strain it can be qualitatively applied to the cracks induced by cyclic loading. The dependencies of the film stress on the distance from the crack edge calculated using the shear lag solution provided by Hsueh&Yanaka [41] for three different thicknesses of silver films on PEN under applied strain of 1.3% are shown in Figure 7. Assuming that that the next crack can propagate if some critical stress value is reached, the 100 nm thick films will have the shortest crack-to-crack distance and the 800 nm thick film will have the larger distance between cracks, as observed in Figure 4a. The exact determination of this critical stress cannot be accomplished using shear lag model since it is a one-dimensional model and the stress concentration at the tip of a propagating crack is not considered.

Although the saturation crack density is predominantly defined by a films thickness, the crack morphology is still dependent on the film microstructure and grain size. The focused ion beam cross sections of the 800 nm thick printed film and the evaporated film with the same thickness are shown in Figure 8. It is clearly seen that the printed film exhibits a brittle crack running perpendicularly to the surface without local thinning or significant plastic deformation. The evaporated film shows the extrusion-crack couple with the voids at the interface indicating that accumulation of dislocation based slip is responsible for the crack formation.

As a final remark one should notice that the evaporated copper films demonstrate electrical and deformation behavior similar to evaporated silver of the same thickness and grain size which confirms that these two length scale effects play a dominant role in the determination of mechanical reliability. This could indicates that the material (i.e. Cu or Ag) used is secondary to the microstructure and thickness in terms of reliability.

## 5. Summary

It was shown that different mechanisms are behind the electro-mechanical reliability determined by monotonic tensile and cyclic bending tests. In the case of monotonic tensile straining, the COS is predominantly determined by the ductility of the film which, depends more on the grain size. The fact that 400 nm thick evaporated silver lines, 800 nm thick evaporated silver lines, and 800 nm thick evaporated copper lines all have similar COS values and similar grain sizes of several hundred nanometers support this finding. The nanocrystalline printed lines demonstrate low COS and significant cracking at 20% strain resulting in low crack densities. The 100 nm thick evaporated Ag lines represent an intermediate case showing relatively low COS, high crack density, but moderate resistance increase at 20% strain. Monotonic tensile straining can be used to determine how far film-substrate couples can be stretched without cracking, which provides information about the ductility of the film.

The cyclic bending reliability depends more on film thickness and the applied strain. When the applied cyclic strain is high enough to for crack initiation and propagation, then the film thickness is the dominant parameter defining the resulting crack density. This explains why the crack density of the evaporated and printed films with the same thickness are very similar despite the differences in the microstructure. At low cyclic strains, however, the value of the minimum strain to activate a dislocation source comes into play. In nanocrystalline materials the dislocation motion is restricted due to the small grain or particle size. For cyclic strains of 0.4% the printed films remain purely elastic and cracks did not form after $5\times10^4$ cycles. In the evaporated films of the same thickness dislocation slip is activated at much lower strains and accounts for the non-zero crack density at the cyclic strain of 0.33%. Among the considered metallization lines the best combination of cyclic and monotonic reliabilities was demonstrated by the inkjet printed silver on PEN, which has a relatively high crack onset strain of 5% as well as zero crack density for cyclic strains below 0.4%.

## 6. Acknowledgements

This work was partially supported by the Austrian Research Promotion Agency (FFG) through the program "Produktion der Zukunft", Project 843648 and the Austrian Science Fund (FWF) through project P27432-N20.

**Tables**

| Table I: Summary of samples and initial characteristics | | | | | |
|---|---|---|---|---|---|
| Sample Name | Substrate & Thickness (μm) | Method & Metal | Thickness (nm) | Initial Resistance (Ohm) | Grain or Particle Size (nm) |
| eTeAg100 | PEN (Teonex), 130 | evap. Ag | 100 | 25 | 100-250 |
| eTeAg400 | PEN (Teonex), 130 | evap. Ag | 400 | 5 | 400-600 |
| eTeAg800 | PEN (Teonex), 130 | evap. Ag | 800 | 2.2 | 500-800 |
| pTeAg800 | PEN (Teonex), 130 | printed Ag | 700-800 | 25 | 30-80 |
| pMeAg800 | PET (Melinex), 180 | printed Ag | 700-800 | 35 | 30-80 |
| eMeCu800 | PET (Melinex), 180 | evap. Cu | 800 | 2.2 | 700-900 |

**Figures & Captions**

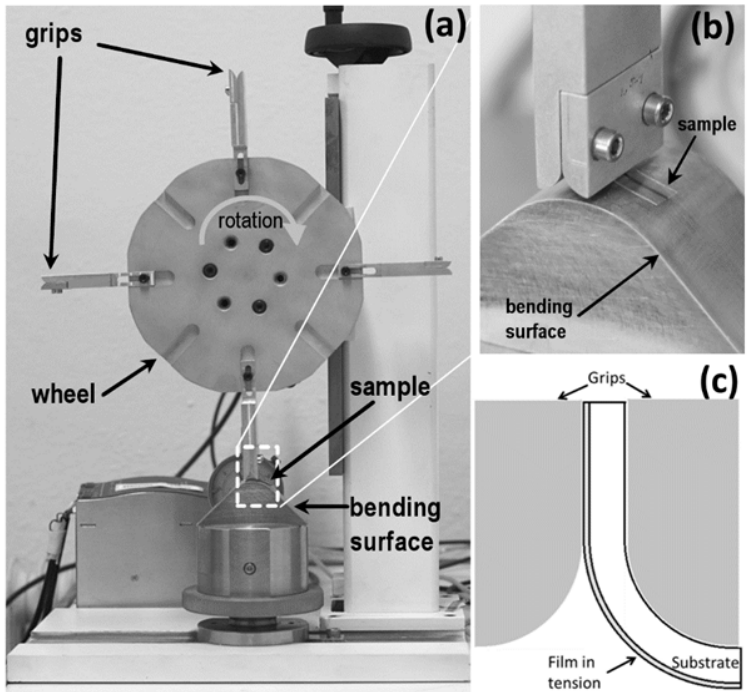

Figure 1: Photograph of the of FLEX-E-TEST bending wheel (a). The sample is clamped between curved grips with a defined radius so that when the sample comes into contact with the bending surface, a defined bending strain is applied (b). When the film is on the outside of the grip, a tensile bending strain is applied (c).

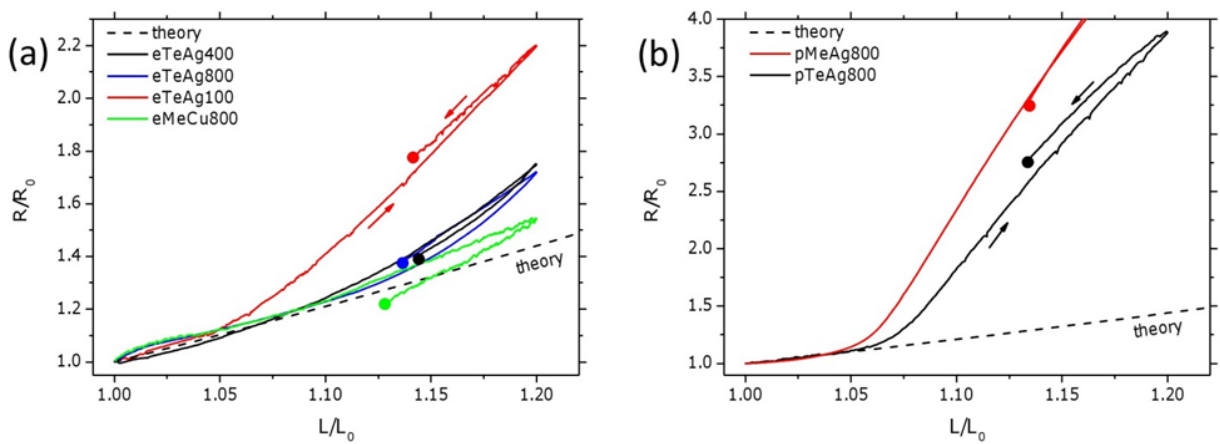

Figure 2: Representative relative electrical resistance measured in-situ during straining as a function of the relative gauge length for evaporated lines (a) and printed lines (b). The arrows

indicate the loading and unloading portions. Solid circles depict the values of resistance and strain at zero force. The dashed theory line represents the constant volume approximation.

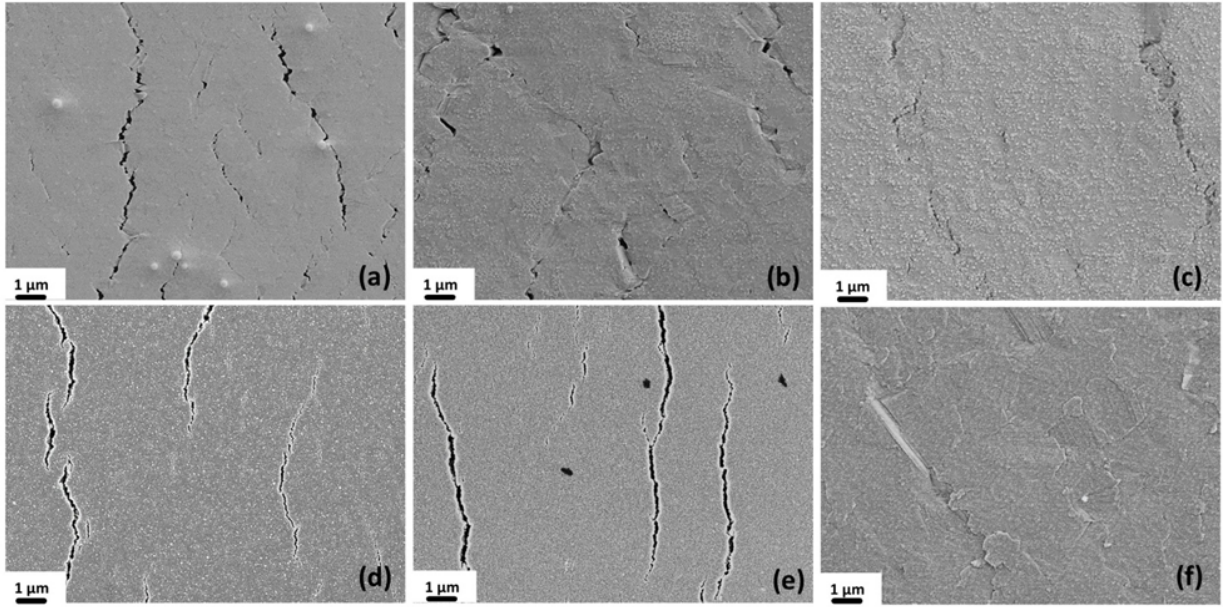

Figure 3: Typical post-mortem SEM micrographs showing the surface of (a) 100 nm thick evaporated Ag, (b) 400 nm thick evaporated Ag, (c) 800 nm thick evaporated Ag, (d) 800 nm thick printed Ag on PEN, (e) 800 nm printed Ag on PET, and (f) 800 nm evaporated Cu after monotonic tensile straining to 20%.

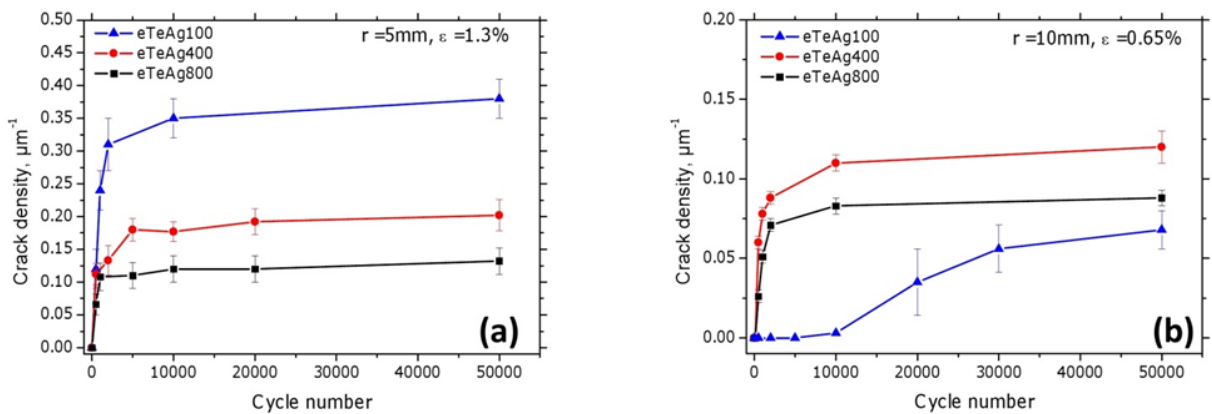

Figure 4: The dependence of the linear crack density on the number of tensile bending cycles for evaporated lines with three different thicknesses and bending radii of (a) 5 mm and (b) 10 mm.

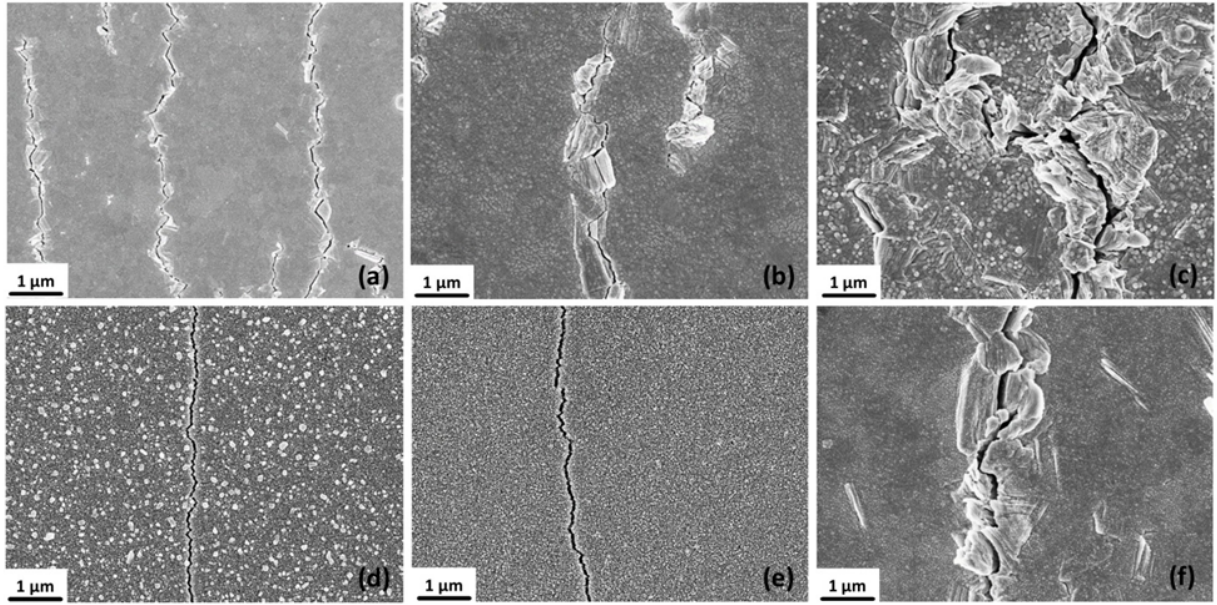

Figure 5: Typical post-mortem SEM micrographs showing the surface of (a) 100 nm thick evaporated Ag, (b) 400 nm thick evaporated Ag, (c) 800 nm thick evaporated Ag, (d) 800 nm thick printed Ag on PEN, (e) 800 nm thick printed Ag on PET, and (f) 800 nm evaporated thick Cu on PET after $5 \times 10^4$ tensile bending cycles with a bending strain of 1.3% (a-d) and 0.9% (e, f).

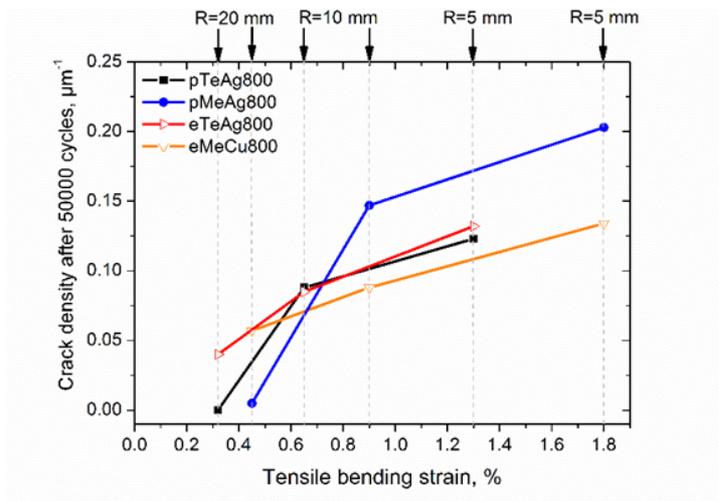

Figure 6: Cyclic reliability curves showing the dependence of the crack density after $5 \times 10^4$ cycles on the applied bending strain. Two evaporated and two printed lines with the same thickness of 800 nm on two different substrates are shown (see also Table I).

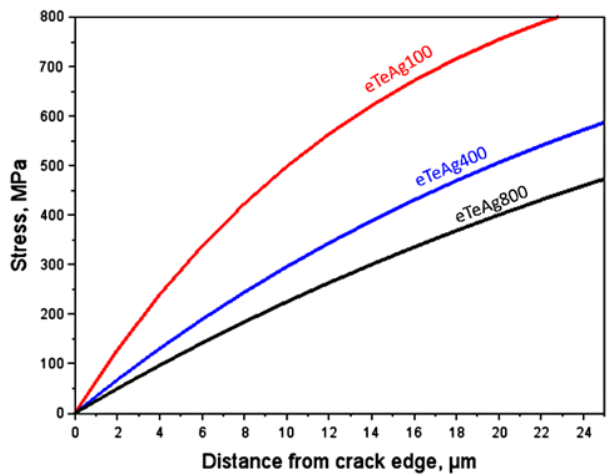

Figure 7: Film stress vs. the distance from the crack edge for three silver films with different thicknesses obtained from shear lag model solution taken from [41].

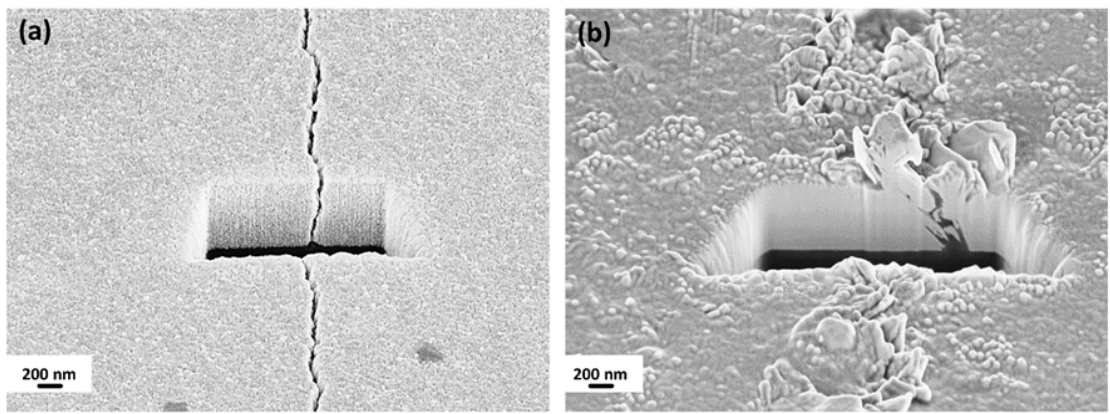

Figure 8: Focused ion beam cross-sections of the cracks induced by cyclic bending in 800 nm printed Ag film (a) and 800 nm thick evaporated Ag film (b).